\documentclass[a4paper]{jpconf}%
\usepackage{amsmath}
\usepackage{graphicx}
\usepackage{amsfonts}
\usepackage{amssymb}%
\begin{document}

\title{Gravity's Rainbow and Black Hole Entropy}
\author{Remo Garattini}

\begin{abstract}
We consider the effects of Gravity's Rainbow on the computation of black hole
entropy using a dynamical brick wall model. An explicit dependence of the
radial coordinate approaching the horizon is proposed to analyze the behavior
of the divergence. We find that, due to the modification of the density of
states, the brick wall can be eliminated. The calculation is extended to
include rotations and in particular to a Kerr black hole in a comoving frame.

\end{abstract}

\address{Universit\`{a} degli Studi di Bergamo, Facolt\`{a} di Ingegneria,
\\Viale Marconi 5, 24044 Dalmine (Bergamo) ITALY.
\\INFN - sezione di Milano, Via Celoria 16, Milan, Italy.}

\ead{remo.garattini@unibg.it}

\section{Introduction}

\label{p1}In recent years many attempts to modify gravity at the fundamental
level have been proposed. Some of them, like $f\left(  R\right)  $
theories\cite{SCMdL}, have been considered to basically change the large scale
structure of the space-time and some others have been conceived to modify the
short scale behavior. Among them, Gravity's Rainbow (GRw) seems to be
promising in dealing with Ultra-Violet divergences (UV). Indeed, in a series
of papers GRw has been used to avoid any regularization/renormalization scheme
which appear in conventional Quantum Field Theory calculations like one loop
corrections to classical quantities\cite{Remo}. This amazing property has been
applied also to black holes and in particular to the computation of black hole
entropy\cite{RemoPLB}. In this last case, the idea is to avoid to introduce a
cut-off of Planckian size known as \textquotedblleft\textit{brick
wall}\textquotedblright\cite{tHooft}. The \textquotedblleft\textit{brick
wall}\textquotedblright\ appears when one uses a statistical mechanical
approach to explain the famous Bekenstein-Hawking
formula\cite{Bekenstein,Hawking}%
\begin{equation}
S_{BH}=\frac{1}{4}A/l_{P}^{2},
\end{equation}
relating the entropy of a black hole and its area. Indeed, when one tries to
adopt such an approach, one realizes that the density of energy levels of
single-particle excitations is divergent near the horizon. Of course, several
attempts have been done to avoid the introduction of the \textit{brick wall}.
For instance, without modifying gravity at any scale, it has been suggested
that the \textit{brick wall} could be absorbed in a renormalization of
Newton's constant\cite{SusUgl,BarEmp,EWin}, while other authors approached the
problem of the divergent brick wall using Pauli-Villars
regularization\cite{DLM,FS,KKSY}. Other than GRw other proposals have been
made in the context of modified gravity. For instance, non-commutative
geometry introduces a natural thickness of the horizon replacing the 't
Hooft's brick wall\cite{BaiYan} and \textit{Generalized Uncertainty Principle}
(GUP) modifies the Liouville measure\cite{Xiang Li,RenQinChun,GAC}. To
understand how GRw works we need to define two unknown functions $g_{1}\left(
E/E_{P}\right)  $ and $g_{2}\left(  E/E_{P}\right)  $ having the following
property%
\begin{equation}
\lim_{E/E_{P}\rightarrow0}g_{1}\left(  E/E_{P}\right)  =1\qquad\text{and}%
\qquad\lim_{E/E_{P}\rightarrow0}g_{2}\left(  E/E_{P}\right)  =1.\label{prop}%
\end{equation}
In this formalism introduced by Magueijo and Smolin\cite{MagSmo}, the
Einstein's field equations are replaced by a one parameter family of equations%
\begin{equation}
G_{\mu\nu}\left(  E\right)  =8\pi G\left(  E\right)  T_{\mu\nu}\left(
E\right)  +g_{\mu\nu}\Lambda\left(  E\right)  ,\label{EEM}%
\end{equation}
where $G\left(  E\right)  $ is an energy dependent Newton's constant and
$\Lambda\left(  E\right)  $ is an energy dependent cosmological constant,
respectively. They are defined so that $G\left(  0\right)  $ is the physical
Newton's constant and $\Lambda\left(  0\right)  $ is the usual cosmological
constant. In this context, the Schwarzschild solution of $\left(
\ref{EEM}\right)  $ becomes%
\begin{equation}
ds^{2}\left(  E\right)  =-\left(  1-\frac{2MG\left(  0\right)  }{r}\right)
\frac{dt^{2}}{g_{1}^{2}\left(  E/E_{P}\right)  }+\frac{dr^{2}}{\left(
1-\frac{2MG\left(  0\right)  }{r}\right)  g_{2}^{2}\left(  E/E_{P}\right)
}+\frac{r^{2}}{g_{2}^{2}\left(  E/E_{P}\right)  }\left(  d\theta^{2}+\sin
^{2}\theta d\phi^{2}\right)  \label{line2}%
\end{equation}
and it can be easily generalized in the following way%
\begin{equation}
ds^{2}\left(  E\right)  =-\exp\left(  -2A\left(  r\right)  \right)  \left(
1-\frac{b\left(  r\right)  }{r}\right)  \frac{dt^{2}}{g_{1}^{2}\left(
E/E_{P}\right)  }+\frac{dr^{2}}{\left(  1-\frac{b\left(  r\right)  }%
{r}\right)  g_{2}^{2}\left(  E/E_{P}\right)  }+\frac{r^{2}}{g_{2}^{2}\left(
E/E_{P}\right)  }d\Omega^{2}.\label{e31}%
\end{equation}
The function $b\left(  r\right)  $ will be referred to as the
\textquotedblleft shape function\textquotedblright\ and it may be thought of
as specifying the shape of the spatial slices. The location of the horizon is
determined by the equation $b\left(  r_{H}\right)  =r_{H}$. On the other hand,
$A\left(  r\right)  $ will be referred to as the \textquotedblleft redshift
function\textquotedblright\ and describes how far the total gravitational
redshift deviates from that implied by the shape function. The line element
$\left(  \ref{e31}\right)  $ describes any spherically symmetric space-time.
It is interesting to wonder what happens when one introduces rotations.
Rotating black holes have a good description in terms of the Kerr metric
which, in the context of GRw, becomes\cite{ZL}%
\begin{equation}
ds^{2}\left(  E\right)  =\frac{g_{tt}dt^{2}}{g_{1}^{2}\left(  E/E_{P}\right)
}+\frac{2g_{t\phi}dtd\phi}{g_{1}\left(  E/E_{P}\right)  g_{2}\left(
E/E_{P}\right)  }+\frac{g_{\phi\phi}d\phi^{2}}{g_{2}^{2}\left(  E/E_{P}%
\right)  }+\frac{g_{rr}dr^{2}}{g_{2}^{2}\left(  E/E_{P}\right)  }%
+\frac{g_{\theta\theta}d\theta^{2}}{g_{2}^{2}\left(  E/E_{P}\right)
},\label{e32}%
\end{equation}
where%
\begin{align}
g_{tt} &  =-\frac{\Delta-a^{2}\sin^{2}\theta}{\Sigma},\qquad g_{t\phi}%
=-\frac{a\sin^{2}\theta\left(  r^{2}+a^{2}-\Delta\right)  }{\Sigma
},\nonumber\\
g_{\phi\phi} &  =\frac{\left(  r^{2}+a^{2}\right)  ^{2}-\Delta a^{2}\sin
^{2}\theta}{\Sigma}\sin^{2}\theta,\qquad g_{rr}=\frac{\Sigma}{\Delta},\qquad
g_{\theta\theta}=\Sigma,
\end{align}
and%
\begin{equation}
\Delta=r^{2}-2MGr+a^{2},\qquad\Sigma=r^{2}+a^{2}\cos^{2}\theta.
\end{equation}
Here $M$ and $a$ are mass and angular momentum per unit mass of the black
hole, respectively. $\Delta$ vanishes when $r=r_{\pm}=MG\pm\sqrt{\left(
MG\right)  ^{2}-a^{2}}$, while $g_{tt}$ vanishes when $r=r_{S\pm}=$
$MG\pm\sqrt{\left(  MG\right)  ^{2}-a^{2}\cos^{2}\theta}$: they are not
modified by GRw and the outer horizon or simply horizon is located at
$r_{+}=r_{H}$. Note that the Kerr metric modified by GRw $\left(
\ref{e32}\right)  $ reduces to the standard rotating black hole background
when $E/E_{P}\rightarrow0$. In this contribution we will consider the effect
of GRw on Black Hole entropy computation even for a rotating background. Units
in which $\hbar=c=k=1$ are used throughout the paper.

\section{GRw Entropy for a Schwarzschild Black Hole}

\label{p2}To see what happens in practice for the Schwarzschild Black Hole, we
define a real massless scalar field whose Euler-Lagrange equations are%
\begin{equation}
\frac{1}{\sqrt{-g}}\partial_{\mu}\left(  \sqrt{-g}g^{\mu\nu}\partial_{\nu
}\right)  \phi=0.\label{EL}%
\end{equation}
The formalism has been outlined in detail in \cite{RemoPLB} and therefore we
refer the reader to \cite{RemoPLB} for details. If $\phi$ has the separable
form
\begin{equation}
\phi\left(  t,r,\theta,\varphi\right)  =\exp\left(  -iEt\right)  Y_{lm}%
(\theta,\varphi)f\left(  r\right)  ,\label{sol}%
\end{equation}
then the equation for $f\left(  r\right)  $ reads%
\begin{equation}
\left[  \frac{g_{2}^{2}\left(  E/E_{P}\right)  \exp\left(  A\left(  r\right)
\right)  }{r^{2}}\partial_{r}\left(  r^{2}\exp\left(  -A\left(  r\right)
\right)  \left(  1-\frac{b\left(  r\right)  }{r}\right)  \partial_{r}\right)
-V_{l}\left(  r\right)  \right]  f_{nl}=0,\label{e33}%
\end{equation}
where%
\begin{equation}
V_{l}\left(  r\right)  =\left(  \frac{l(l+1)}{r^{2}}-\frac{E_{nl}^{2}g_{1}%
^{2}\left(  E/E_{P}\right)  \exp\left(  2A\left(  r\right)  \right)  }%
{1-\frac{b\left(  r\right)  }{r}}\right)
\end{equation}
and where $Y_{lm}(\theta,\varphi)$ is the usual spherical harmonic function.
In order to use the WKB approximation to compute the entropy, we define the
following r-dependent radial wave number $k(r,l,E)$
\begin{equation}
k_{r}^{2}(r,l,E)\equiv\frac{1}{\left(  1-\frac{b\left(  r\right)  }{r}\right)
}\left[  \exp\left(  2A\left(  r\right)  \right)  \frac{E^{2}h^{2}\left(
E/E_{P}\right)  }{\left(  1-\frac{b\left(  r\right)  }{r}\right)  }%
-\frac{l(l+1)}{r^{2}}\right]  ,\label{squarekr}%
\end{equation}
where%
\begin{equation}
h\left(  E/E_{P}\right)  =\frac{g_{1}\left(  E/E_{P}\right)  }{g_{2}\left(
E/E_{P}\right)  }.\label{h(E)}%
\end{equation}
The number of modes with frequency less than $E$ is given approximately by
\begin{equation}
n(E)=\frac{1}{\pi}\int_{0}^{l_{max}}(2l+1)\int_{r_{H}}^{R}\sqrt{k^{2}%
(r,l,E)}drdl,
\end{equation}
Here it is understood that the integration with respect to $r$ and $l$ is
taken over those values which satisfy $r_{H}\leq r\leq R$ and $k^{2}%
(r,l,E)\geq0$. Thus, from Eq.$\left(  \ref{squarekr}\right)  $ we get%
\begin{equation}
\frac{dn(E)}{dE}=\frac{2}{\pi}\frac{d}{dE}\left(  \frac{1}{3}E^{3}h^{3}\left(
E/E_{P}\right)  \right)  \int_{r_{H}}^{R}dr\frac{\exp\left(  3A\left(
r\right)  \right)  }{\left(  1-\frac{b\left(  r\right)  }{r}\right)  ^{2}%
}r^{2}.\label{states}%
\end{equation}
Since the free energy can be written as
\begin{equation}
F=\frac{1}{\beta}\int_{0}^{\infty}\ln\left(  1-e^{-\beta E}\right)
\frac{dn(E)}{dE}dE,\label{F0}%
\end{equation}
where $\beta$ is the inverse temperature measured at infinity. Plugging
Eq.$\left(  \ref{states}\right)  $ into $\left(  \ref{F0}\right)  $ we find
\begin{equation}
F_{r_{H}}=\frac{2}{\pi}\frac{1}{\beta}\int_{0}^{\infty}\ln\left(  1-e^{-\beta
E}\right)  \frac{d}{dE}\left(  \frac{1}{3}E^{3}h^{3}\left(  E/E_{P}\right)
\right)  \left[  \int_{r_{H}}^{r_{1}}drr^{2}\frac{\exp\left(  3A\left(
r\right)  \right)  }{\left(  1-\frac{b\left(  r\right)  }{r}\right)  ^{2}%
}\right]  dE
\end{equation}
and%
\begin{equation}
F_{R}=\frac{2}{\pi}\frac{1}{\beta}\int_{0}^{\infty}\ln\left(  1-e^{-\beta
E}\right)  \frac{d}{dE}\left(  \frac{1}{3}E^{3}h^{3}\left(  E/E_{P}\right)
\right)  \left[  \int_{r_{1}}^{R}drr^{2}\frac{\exp\left(  3A\left(  r\right)
\right)  }{\left(  1-\frac{b\left(  r\right)  }{r}\right)  ^{2}}\right]  dE.
\end{equation}
Assuming that $A\left(  r\right)  <\infty,$ $\forall r\in\left[  r_{H}%
,+\infty\right)  $, $F_{R}$ is dominated by large volume effects for large $R$
and it will give the contribution to the entropy of a homogeneous quantum gas
in flat space at a uniform temperature $T$ when GRw is considered. If GRw does
not come into play, then the radial part of $F_{r_{H}}$ becomes divergent in
proximity of $r_{H}$. On the other hand, if we allow that the
\textquotedblleft\textit{brick wall }$r_{0}$\textquotedblright\ be affected by
GRw, namely $r_{0}\equiv r_{0}\left(  E/E_{P}\right)  $, then the radial
integration $F_{r_{H}}$ becomes%
\begin{gather}
\int_{r_{H}+r_{0}\left(  E/E_{P}\right)  }^{r_{1}}drr^{2}\frac{\exp\left(
3\Lambda\left(  r\right)  \right)  }{\left(  1-\frac{b\left(  r\right)  }%
{r}\right)  ^{2}}\simeq r_{H}^{4}\frac{\exp\left(  3A\left(  r_{H}\right)
\right)  }{\left(  1-b^{\prime}\left(  r_{H}\right)  \right)  ^{2}}\frac
{1}{r_{0}\left(  E/E_{P}\right)  }\nonumber\\
\nonumber\\
=r_{H}^{3}\frac{\exp\left(  3A\left(  r_{H}\right)  \right)  }{\left(
1-b^{\prime}\left(  r_{H}\right)  \right)  ^{2}}\frac{1}{\sigma\left(
E/E_{P}\right)  },\label{rh}%
\end{gather}
where we have assumed that, in proximity of the throat the brick wall can be
written as $r_{0}\left(  E/E_{P}\right)  =r_{H}\sigma\left(  E/E_{P}\right)  $
with%
\begin{equation}
\sigma\left(  E/E_{P}\right)  \rightarrow0,\qquad E/E_{P}\rightarrow0.
\end{equation}
Plugging Eq.$\left(  \ref{rh}\right)  $ into $F_{r_{H}}$ we obtain, after an
integration by parts%
\begin{equation}
F_{r_{H}}=-\frac{C_{r_{H}}}{3\beta r_{H}}\int_{0}^{\infty}\frac{E^{3}%
h^{3}\left(  E/E_{P}\right)  }{\sigma\left(  E/E_{P}\right)  }\left[
\frac{\beta}{\left(  \exp\left(  \beta E\right)  -1\right)  }-\frac{\ln\left(
1-e^{-\beta E}\right)  }{E_{P}\sigma\left(  E/E_{P}\right)  }\sigma^{\prime
}\left(  E/E_{P}\right)  \right]  .\label{Frw}%
\end{equation}
$h\left(  E/E_{P}\right)  $ must be chosen in such a way to allow the
convergence when $E/E_{P}\rightarrow\infty$, thus we assume that%
\begin{equation}
h\left(  E/E_{P}\right)  =\exp\left(  -\frac{E}{E_{P}}\right)  \qquad
\mathrm{and}\qquad\sigma\left(  E/E_{P}\right)  =h^{\delta}\left(
E/E_{P}\right)  \left(  \frac{E}{E_{P}}\right)  ^{\alpha}.\label{choice}%
\end{equation}
In particular, for $\delta=0;$ $\alpha=2$, one finds that the entropy becomes%
\begin{equation}
S=\beta^{2}\frac{\partial F_{r_{H}}}{\partial\beta}\underset{\beta E_{P}\gg
1}{\simeq}\frac{A_{r_{H}}E_{P}^{2}}{4}\frac{\exp\left(  2A\left(
r_{H}\right)  \right)  }{1-b^{\prime}\left(  r_{H}\right)  }\frac{2}{9\pi}.
\end{equation}
where we have used the expression for the surface gravity in the low energy
limit. As we can see the \textquotedblleft\textit{brick wall}%
\textquotedblright\ does not appear.

\section{GRw Entropy for the Kerr Black Hole}

To discuss the entropy for a Kerr black hole we have two options: we can use a
rest observer at infinity (ROI) or we can use a Zero Angular Momentum Observer
(ZAMO)\cite{CYLY, Lee Kim}. The ROI\ frame is described by the line element
$\left(  \ref{e32}\right)  $ and the appropriate form of the free energy is
the following%
\begin{equation}
F=\frac{1}{\beta}\int_{0}^{\infty}dn\left(  E\right)  \ln\left(
1-e^{-\beta\left(  E-m\Omega\right)  }\right)  .\label{FK}%
\end{equation}
It is immediate to see that when we use a ROI, the problem of superradiance
appears when the free energy $\left(  \ref{FK}\right)  $ is computed in the
range $0<E<m\Omega$. On the other hand when a ZAMO is considered, the free
energy $\left(  \ref{FK}\right)  $ becomes similar to the one used for a
Schwarzschild black hole $\left(  \ref{F0}\right)  $. Basically this happens
because near the horizon the metric becomes
\begin{equation}
ds^{2}=-\frac{N^{2}dt^{2}}{g_{1}^{2}\left(  E/E_{P}\right)  }+g_{\phi\phi
}\frac{d\phi^{2}}{g_{2}^{2}\left(  E/E_{P}\right)  }+g_{rr}\frac{dr^{2}}%
{g_{2}^{2}\left(  E/E_{P}\right)  }+g_{\theta\theta}\frac{d\theta^{2}}%
{g_{2}^{2}\left(  E/E_{P}\right)  }\label{e32c}%
\end{equation}
and the mixing between $t$ and $\phi$ disappears. Moreover when we use a ZAMO
frame, the superradiance does not come into play because there is no
ergoregion. Indeed since we have defined%
\begin{equation}
N^{2}=g_{tt}-\frac{g_{t\phi}^{2}}{g_{\phi\phi}}=-\frac{1}{g^{tt}}%
=-\frac{\Delta\sin^{2}\theta}{g_{\phi\phi}},
\end{equation}
$N^{2}$ vanishes when $r\rightarrow r_{H}$. Therefore if we repeat the same
steps which led us to the computation of $\left(  \ref{states}\right)  $, one
finds%
\begin{equation}
\frac{dn(E)}{dE}=\frac{1}{8\pi^{2}}\int d\theta d\bar{\phi}\int_{r_{H}}%
^{R}dr\left(  -g^{tt}\right)  ^{\frac{3}{2}}\sqrt{g_{rr}g_{\theta\theta
}g_{\phi\phi}}\frac{1}{3}\frac{d}{dE}\left(  h^{3}\left(  E/E_{P}\right)
E^{3}\right)  ,
\end{equation}
where the solution of the massless Klein-Gordon equation $\left(
\ref{EL}\right)  $ assumes the form%
\begin{equation}
\phi\left(  x\right)  =\exp\left(  -iEt+im+iK\left(  r,\theta\right)  \right)
\end{equation}
with%
\begin{equation}
k_{r}=\frac{\partial K\left(  r,\theta\right)  }{\partial r},\qquad k_{\theta
}=\frac{\partial K\left(  r,\theta\right)  }{\partial\theta}%
\end{equation}
defined in such a way to use the WKB approximation. In proximity of the
horizon, the free energy can be approximated by%
\begin{equation}
F_{r_{H}}=\frac{1}{8\pi^{2}\beta}\int d\theta d\bar{\phi}\int_{0}^{\infty}%
\ln\left(  1-e^{-\beta E}\right)  \frac{d}{dE}\left(  \frac{1}{3}h^{3}\left(
E/E_{P}\right)  E^{3}\right)  dE\int_{r_{H}}^{r_{1}}dr\left(  -g^{tt}\right)
^{\frac{3}{2}}\sqrt{g_{rr}g_{\theta\theta}g_{\phi\phi}}%
\end{equation}
which can be further reduced to%
\begin{equation}
F_{r_{H}}\simeq\frac{C\left(  r_{H},\theta\right)  }{8\pi^{2}\beta}\int
_{0}^{\infty}\frac{\ln\left(  1-e^{-\beta E}\right)  }{\sigma\left(
E/E_{P}\right)  }\frac{d}{dE}\left(  \frac{1}{3}h^{3}\left(  E/E_{P}\right)
E^{3}\right)  dE,
\end{equation}
where%
\begin{equation}
C\left(  r_{H},\theta\right)  =\int d\theta d\bar{\phi}\left[  \frac{\left(
r_{H}^{2}+a^{2}\right)  ^{4}\sin\theta}{r_{H}\left(  r_{H}-r_{-}\right)
^{2}\Sigma_{H}}\right]  .
\end{equation}
With an integration by parts one finds%
\begin{equation}
F_{r_{H}}=-\frac{C\left(  r_{H},\theta\right)  }{24\pi^{2}\beta}\int
_{0}^{\infty}\frac{E^{3}h^{3}\left(  E/E_{P}\right)  }{\sigma\left(
E/E_{P}\right)  }\left[  \frac{\beta}{\left(  \exp\left(  \beta E\right)
-1\right)  }-\frac{\ln\left(  1-e^{-\beta E}\right)  }{E_{P}\sigma\left(
E/E_{P}\right)  }\sigma^{\prime}\left(  E/E_{P}\right)  \right]  dE.
\end{equation}
If we adopt the same choice of the previous section described by $\left(
\ref{choice}\right)  $ and we fix our attention on the particular values
$\delta=0$ and $\alpha=2$, one finds%
\begin{align}
F_{r_{H}} &  =-\frac{C\left(  r_{H},\theta\right)  }{24\pi^{2}\beta}\int
_{0}^{\infty}\left[  \frac{\beta Ee^{-3E/E_{P}}}{\left(  \exp\left(  \beta
E\right)  -1\right)  }-2e^{-3E/E_{P}}\ln\left(  1-e^{-\beta E}\right)
\right]  dE\nonumber\\
&  =-\frac{C\left(  r_{H},\theta\right)  }{24\pi^{2}\beta}\left[  \zeta\left(
2,1+\frac{3}{\beta E_{P}}\right)  +\frac{\beta E_{P}}{3}\left(  \gamma
+\Psi\left(  1+\frac{3}{\beta E_{P}}\right)  \right)  \right]  ,
\end{align}
where $\zeta\left(  s,\nu\right)  $ is the Hurwitz zeta function,
$\Gamma\left(  x\right)  $ is the gamma function and $\Psi\left(  x\right)  $
is the digamma function. In the limit where $\beta E_{P}\gg1$, at the leading
order, one finds that the entropy can be approximated by%
\begin{equation}
S=\beta^{2}\frac{\partial F_{r_{w}}}{\partial\beta}=\frac{E_{P}^{2}}{36\beta
}\int d\theta d\bar{\phi}\left[  \frac{\left(  r_{H}^{2}+a^{2}\right)
^{4}\sin\theta}{r_{H}\left(  r_{H}-r_{-}\right)  ^{2}\Sigma_{H}}\right]
\label{S}%
\end{equation}
and even in this case the \textquotedblleft\textit{brick wall}%
\textquotedblright\ does not appear. Of course the entropy $\left(
\ref{S}\right)  $ can always be cast in the familiar form%
\begin{equation}
S=\frac{A_{H}}{4G},
\end{equation}
where $A_{H}$ is the horizon area. To summarize, we have shown that the
ability of Gravity's Rainbow to keep under control the UV divergences applies
also to rotations. However the connection between a ROI and a ZAMO has to be
investigated with care\cite{Kerr}. Indeed in the ROI frame, the superradiance
phenomenon appears, while in the ZAMO frame does not. Once the connection is
established nothing forbids to extend this result to other rotating
configuration like, for example, Kerr-Newman or Kerr-Newman-De Sitter (Anti-De Sitter).

\section*{Acknowledgments}

The author would like to thank MDPI for a partial financial support.

\end{document}